\documentclass[12pt]{article}

\usepackage{epsf}
\usepackage{pstricks}
\begin{document}
\bigskip

{\hfill \bf \large THE STUDY OF THE X-RAY EMISSION FROM \hfill}

{ \bf \large THE ACCRETING BLACK  HOLES AND NEUTRON STARS }

{\hfill \it \bf PhD thesis of Mikhail Revnivtsev \hfill }

{\hfill \it Advisor Marat Gilfanov \hfill }

\bigskip

{\bf \hfill Summary \hfill}

\bigskip

The thesis studies the X--ray emission from the Galactic compact objects
(accreting neutron stars and black holes), using mainly the RXTE data.

In particular following results have been included: spectral evolution
 of X--ray transients GRS 1739--278, XTE J1755--324, GS 1354--644, XTE
J1748--288; the outburst and detailed spectroscopy of the anomalous X-ray
Novae
XTE J0421+560; study of the outburst of the millisecond pulsar SAX
J1808.4--3658 and stability of its spectrum; study of the SAX
J1808.4--3658 pulse profile and relativistic distortions of the profile, the
linear velocity of the emitting area on the surface of the neutron star SAX
J1808.4--3658, inferred from our simplest model is about $\sim$0.1\,c and the
neutron star radius $\sim$13 km;
frequency resolved spectroscopy and its application to the
hard state of Cyg X-1 and GX~339-4, correlations of the spectral and
timing properties for Cyg X-1, GX~339-4 and possible connection to the
disk-spheroid model of the accreting flow in the vicinity of the black hole.

\bigskip

\bigskip

\sloppypar

{\bf\em The first part} of the thesis briefly describes the RXTE, GRANAT and 
MIR/KVANT/TTM instruments and observational prodecures.

{\bf\em The second part} consists of 4 chapters, which describe the
outbursts of the X-ray Novae.

{\em The first chapter} of the second part describes the outburst of X-ray
Novae GRS 1739--278 in Mar.-May 1996 in details. We used RXTE and
MIR/KVANT/TTM data in our analysis. It is shown, that on the beginning of
the outburst (before the primary maximum) the  spectrum of the source had
very weak soft component , its contribution to the 2--20 keV flux did not
exceed 50\% (TTM telescope result). For a comparison, the usual value for
the soft component contribution for the High State spectra is about
80--90\%. The analysis of the subsequent observations of RXTE showed the
change of the ``inner disk temperature'', that can not be explained by the
standard disk theory with the constant inner radius. We propose, that the
electron scattering plays the important role in the forming the emergent
spectrum in GRS 1739--278.

{\em The second chapter} describes the outburst of XTE J1755--324 in 1997
using RXTE and GRANAT data. The RXTE observations showed the sequence of
standard states of X-ray Novae : High (strong soft component, weak aperiodic 
noise) and Low (hard spectrum and significant chaotic variability). The
peculiar hardening was detected close to the tertiary maximum.

{\em In the third chapter} we explore the outburst of the  hard X-ray Novae
GS 1354--644 in the end of 1997 -- beginning of 1998. It is shown, that the
energy spectrum of this Nova can be described by the emission of the
comptonized cloud with the temperature $\sim20$ keV with reflected
component. The analysis of aperiodic variability of X-ray flux showed, that
it can be represented by the sequence of the exponential shots with two (in
the first two observations - three) different characteristic decay times.
The analysis of the density of a probability of X-ray flux, integrated over
16 sec bins, allowed us to estimate the shot rate value of the long shots.

It was shown that the dependence of the rms of the aperiodic variability on
the photon energy can not be explained by the compton upscattering process.
It was shown that majority of the black hole systems in the hard state
demonstrate the decrease of the rms value with the photon energy, while the
neutron star systems show the opposite trend -- the increase of the rms with 
the photon energy.

\begin{figure}[thb]
\hbox{
\begin{minipage}{10.5cm}
\epsfxsize=10cm
\epsffile{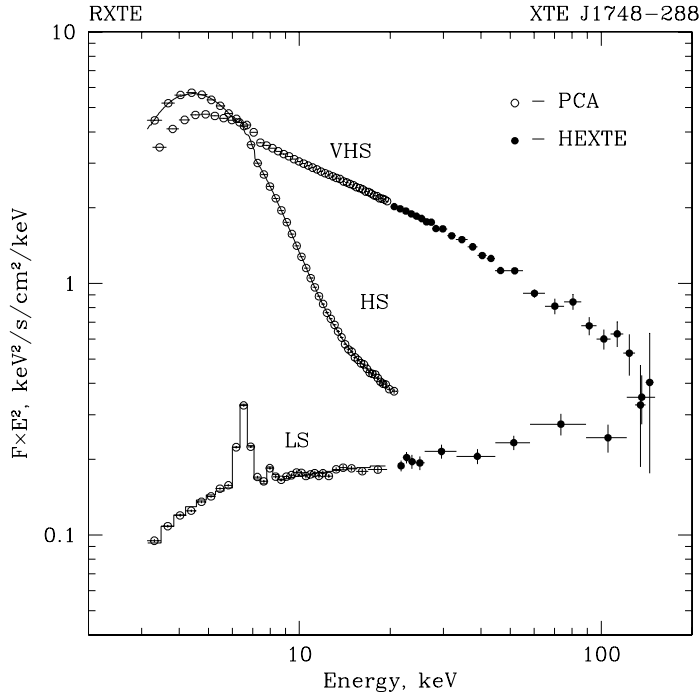}
\end{minipage}
\begin{minipage}{5cm}
\caption{\small
Typical energy spectra of XTE J1748--288 in the different states. The white
and black circles show the PCA and HEXTE data respectively. }
\end{minipage}
}
\end{figure}

{\em The fourth chapter} of the second part describes the outburst of the
third Galactic miqroquasar XTE J1748--288 in 1998. We detected the sequence
of the standard states : Very High, High and Low. During the Very High state 
observations the source had very weak soft component, similar to that of GRS 
1739--278 (see Chapter 1 above).

In {\bf\em the third part} of the thesis we describe the outburst of the 
very peculiar source -- XTE J0421+560/CI Cam in 1998. It is shown that the
emission of the source originates in the hot optically thin plasma cloud.
We show that the source do not demonstrate any aperiodic variability (in the 
first observations the $2\sigma$ upper limit on the rms value is about
0.8\%). The evolution of the source spectrum is described in details.
The mass of the emitting object and its radius are estimated. We detected
the displacement in the position of the broad emission feature (from 6.53
keV to 6.61 keV). we propose that the apparent displacement of the centroid
can be caused by the presence of 6.4 keV fluorescent line in the spectra.
We detected the additional emission feature at the energy $\sim8$ keV (most
likely -- the thermal lines $K_{\beta}$ Fe and $K_{\alpha}$ Ni)

\begin{figure}[thb]
\hbox{
\begin{minipage}{9.5cm}
\epsfxsize=9cm
\epsffile{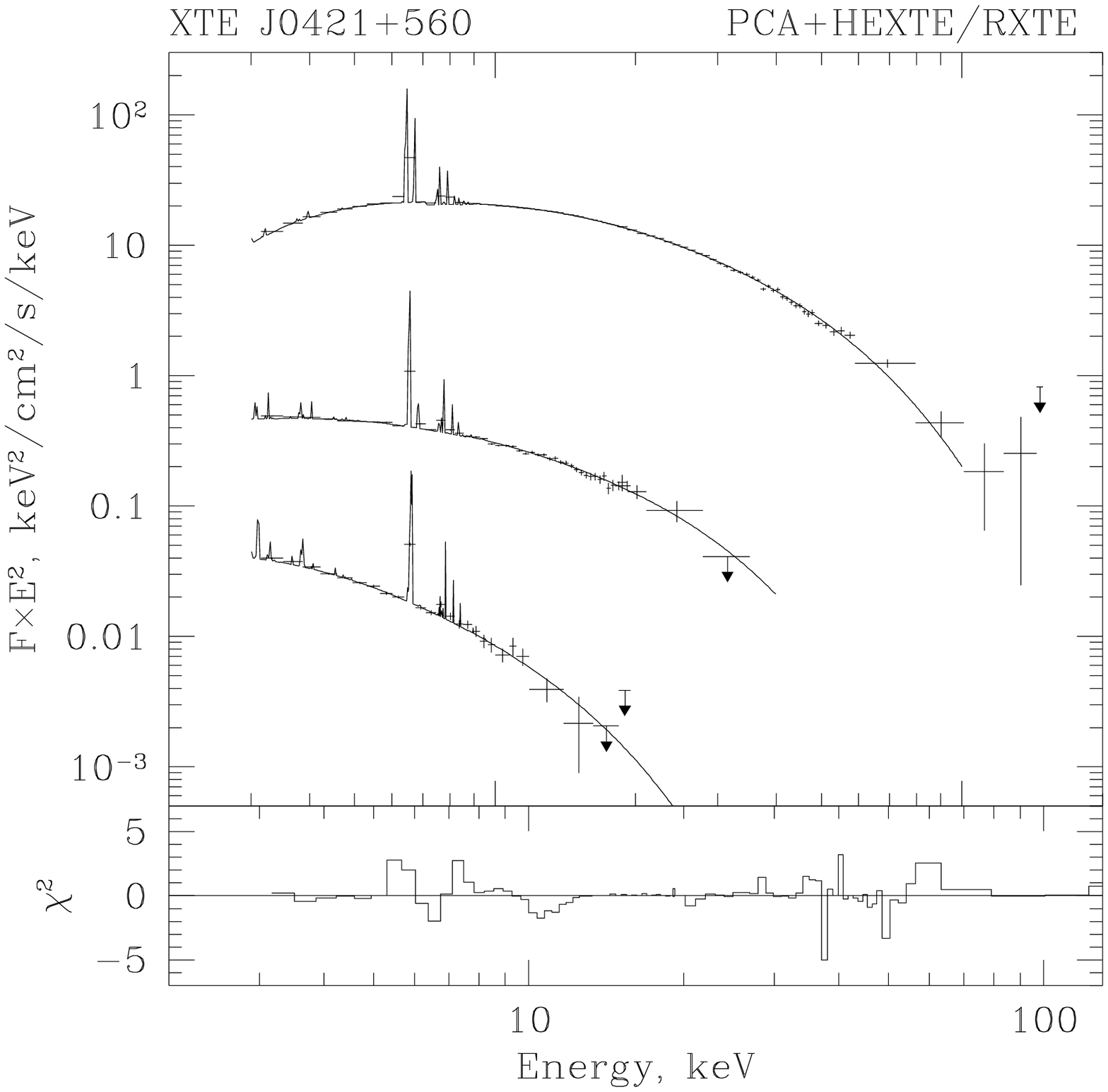}
\end{minipage}
\begin{minipage}{5cm}
\caption{\small
Spectra of XTE J0421+560 in different luminosity states. The
most luminous one -- on Apr. 1, 1998, the middle one -- Apr. 3, 1998, the
weakest -- Apr. 9, 1998. On the lower panel the $\chi^2$ curve for the upper
spectrum is presented. Not all of shown lines are seen by PCA, a number of
emission lines caused by $meka$ model (see text).}
\end{minipage}
}
\end{figure}

\bigskip

In {\bf\em the fourth part} of the work, that consist of two chapters, we
analise millisecond pulsar-burster SAX J1808.4--3658.  

{\em The first chapter} of this part devoted to the outburst of SAX
J1808.4--3658. It is shown that the energy spectrum of the source
($I_\nu\sim\nu^{-2}$) was approximately the same throughout the outburst,
while the energy flux changed by the two orders of magnitude. Assuming that
the rapid drop of the X-ray light curve of the SAX J1808.4--3658 is caused by
transition of the source to a propeller regime we estimated the star
magnetic field.

$$
B\sim 3\cdot 10^7
M_{1.4}^{1/3} R_{6}^{-8/3}
\left(\frac{L_x}{10^{35}}\right)^{1/2} ~\mathrm{G}
$$

 (here $M_{1.4}$ -- the neutron star mass in 1.4$M_{\odot}$, $R_6$ -- the
 radius of the  star in tens of km, $L_x$ -- the transition
 luminosity). However, the rapid drop of the X-ray flux can also be caused
 by the disk instabilities, analogous to that observed in cataclysmic
 variables. We propose, that the hard spectrum can originate in the
 radiation dominated shock near the neutron star surface.

\begin{figure}[htb]
\hbox{
\epsfxsize=6cm
\epsffile[50 100 560 560]{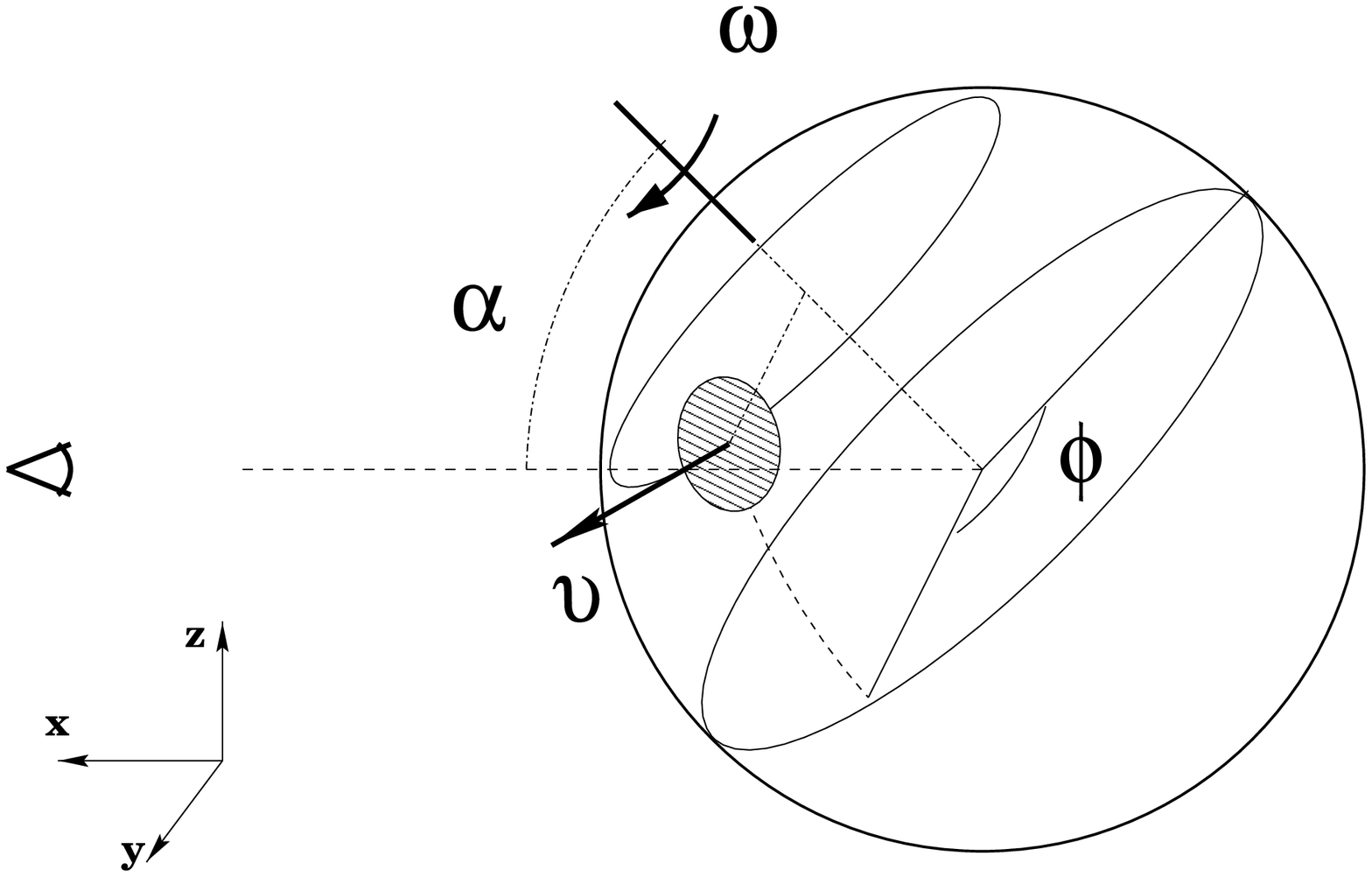}
\epsfxsize=9cm
\epsffile[50 220 590 580]{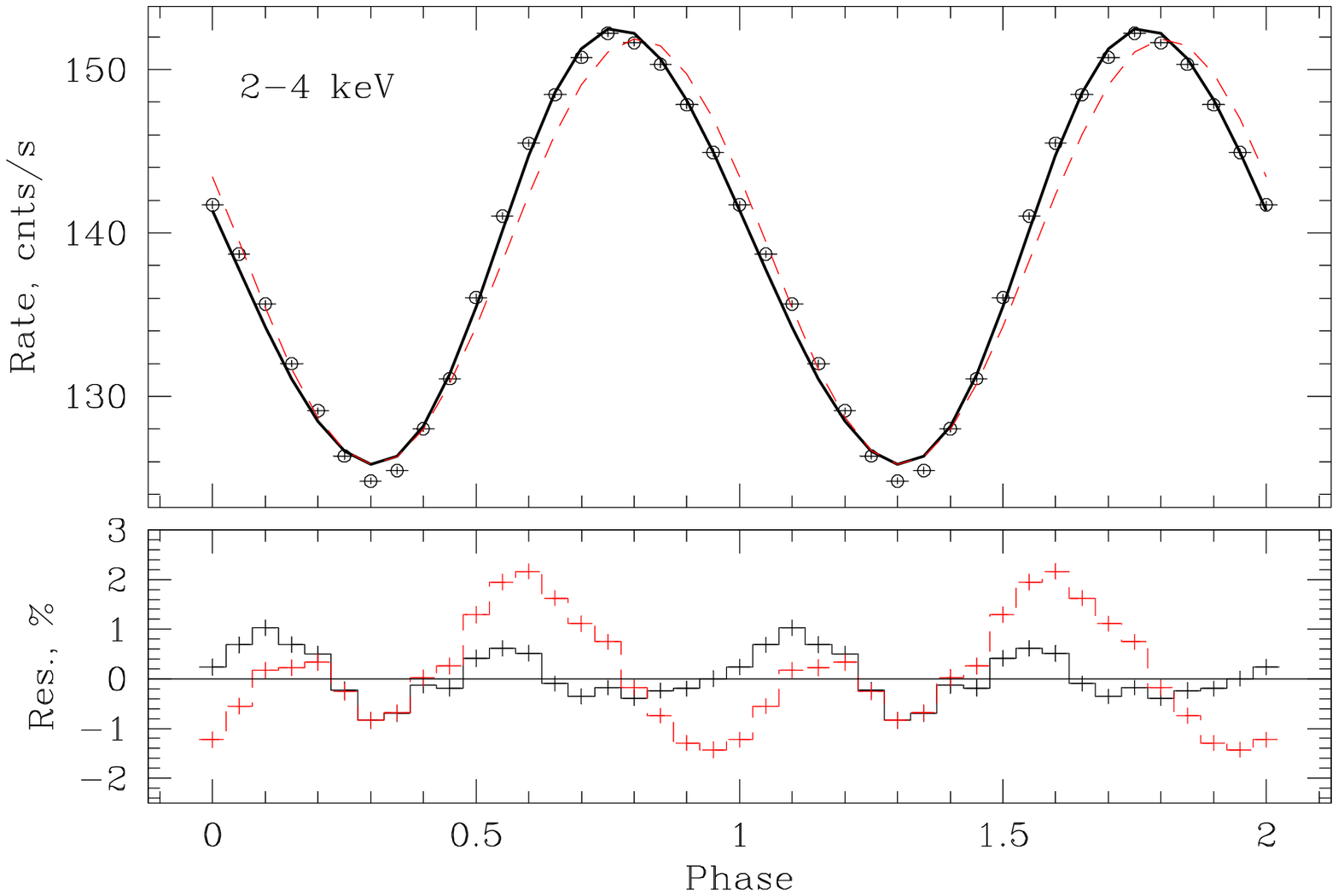}
}
\caption{\small The pulse profile of SAX
J1808.4--3658. The solid line shows the approximation of the  observed pulse 
profile by our model with the relativistic distortions. Dashed line shows
the  approximation by the simple sinus. Lower panel shows the residuals in 
percents. Left picture shows the simplest geometry used in the model.
}

\end{figure}

In the {\em second chapter} of the fourth part we analyze the pulse profile
of the millisecond pulsar SAX
J1808.4--3658. It is shown that the observed difference of the pulse
profile in the energy $<$4 keV (in the energy spectrum at these energies the 
soft component dominates, the soft component can be approximated with the
black body spectrum) from the simple sinusoid can be explained by the
influence of the relativistic abberation (Fig.~3). We propose the simplest
model of the observed distortions. This model allowed us to estimate the
emitting region linear velocity and, consequently, the radius of the
neutron star: $R\sim$13--19 km. It is shown the allowance for Schwarcschild
metric further improve the pulse profile approximation.
  
\bigskip

In {\bf\em the fifth part} we explore the aperiodic variability of the well
known Galactic black hole candidates Cyg X-1 and GX~339-4. We propose the
frequency resolved spectral analysis for the exploration of the variability
of the various spectral component. With the help of this method we show that 
the variability of the reflected component in the spectra of Cyg X-1 and GX
339-4 is strongly dumped at the frequencies higher than 1--10 Hz. We propose 
simplest explanations of this behavior. In particular, the suppression of
the high frequency variability can be caused by the influence of the large
size of the reflecting accretion disk. 

\begin{figure}[tb]
\hbox{
\begin{minipage}{9.5cm}
\epsfxsize=8cm
\epsffile[80 220 560 600]{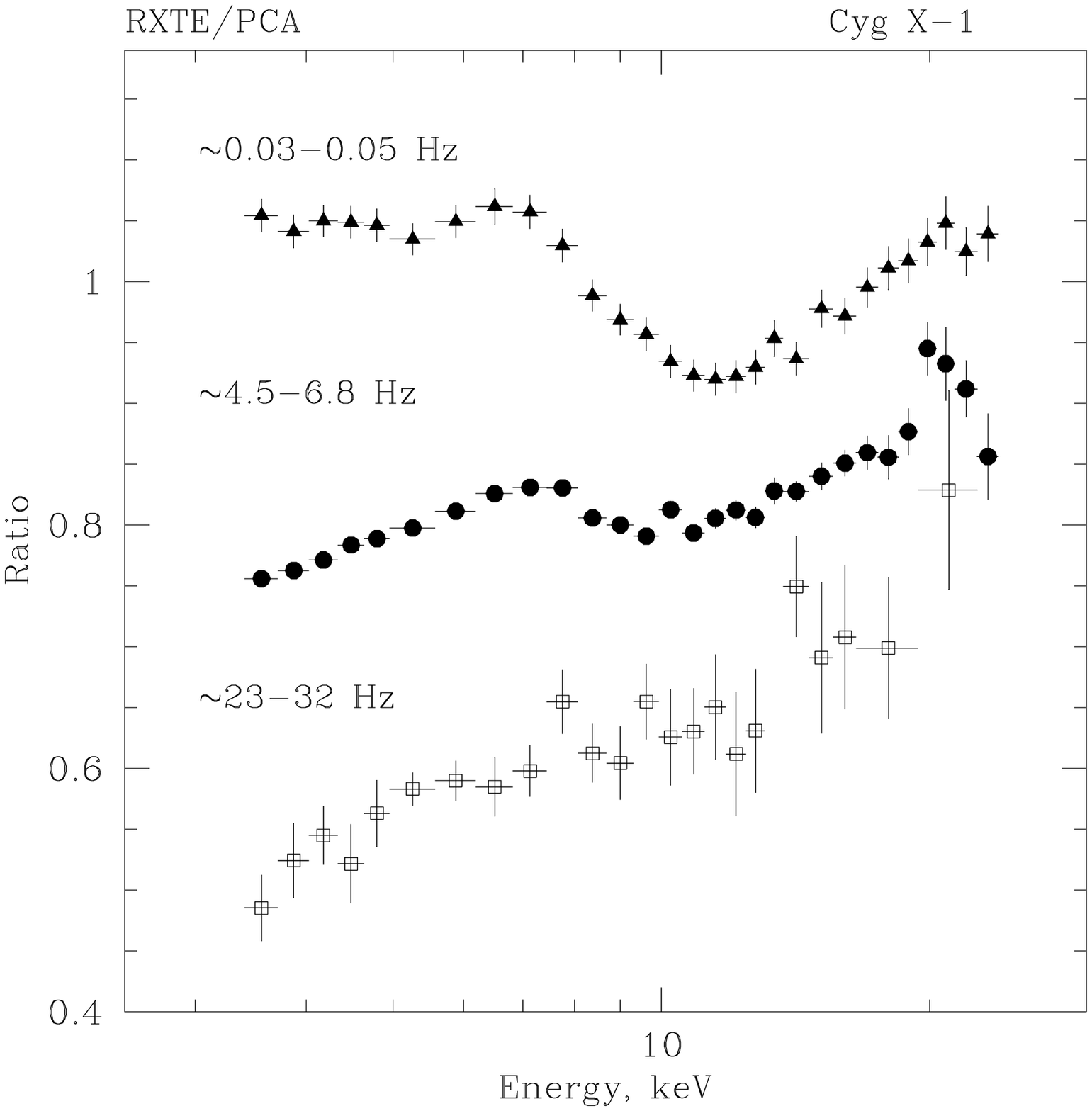}
\end{minipage}
\begin{minipage}{5cm}
\caption{\small The ratio of the energy spectra of Cyg X-1 in different frequency
bands to a power law model with photon index $\alpha=1.8$. Spectra
corresponding to 0.03-0.05 Hz and 23--32 Hz were rescaled for clarity.
}
\end{minipage}
}

\end{figure} 
 
\bigskip

\begin{figure}[htb]
\hbox{
\epsfxsize=7cm
\epsffile[10 110 600 440]{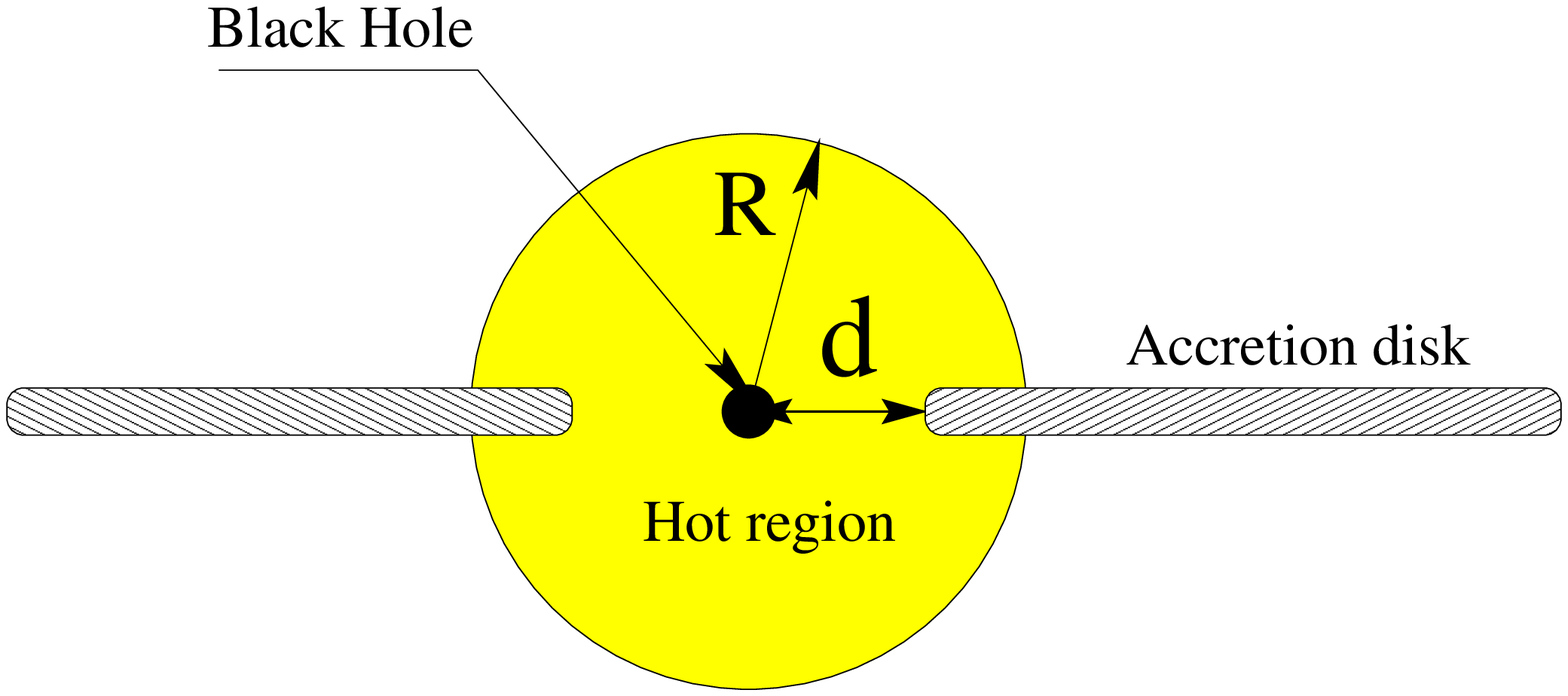}
\epsfxsize=8cm
\epsffile{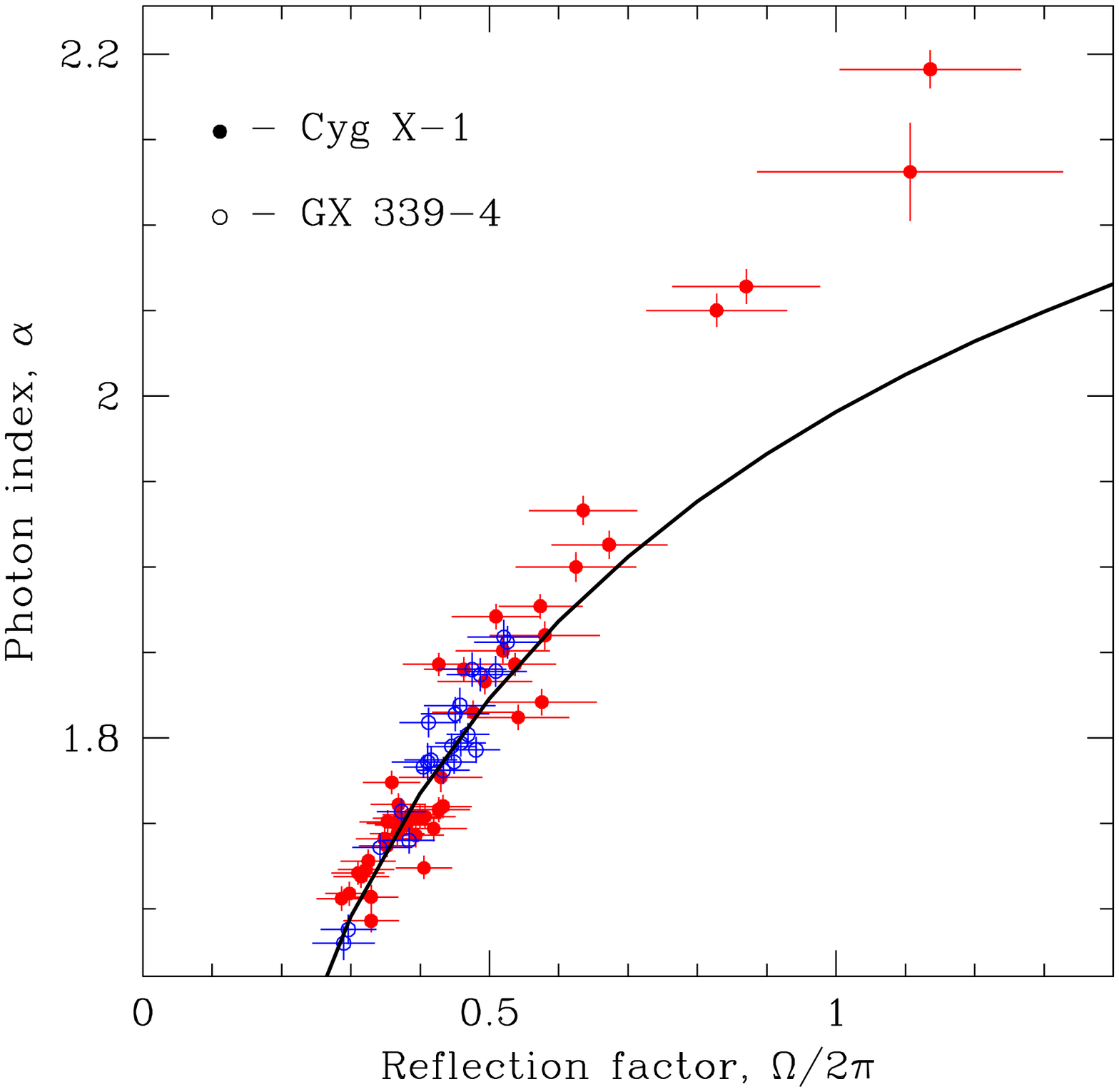}
}
\begin{minipage}{15.5cm}
\caption{\small
 {\bf\em Left panel}: The assumed geometry of the accretion flow in Cyg X-1
 and GX 339-4 in the low spectral state {\bf\em Right panel}: The
 approximation the correlation $\Omega/2\pi-\alpha$ observed for Cyg X-1 and
 GX 339-4. The solid line shows the used simplest model.
}
\end{minipage}
\end{figure} 

{\bf\em The sixth part} of the thesis is devoted to the systematic analysis
of the energy spectra and the characteristics of the  aperiodic variability
of Cyg X-1 and GX~339-4 in the hard/low spectral state.
We detected tight correlations of the parameters of the energy spectra and
the power density spectra. The increase of the characteristic frequency on
the power density spectra (the frequency of the first break or the QPO
frequency) accompanying by the increase of photon index $\alpha$ of the
primary power law  and by the increase of the reflection amplitude
$\Omega /2 \pi$. It is commonly accepted now, that the accretion flow in
the black hole systems in the hard/low spectral state consist of the hot
inner cloud, where the comptonized component originates, and the surrounding 
cold accretion disk (see e.g. Fig.5). The observed correlations can be understood if we
assume that the characteristic frequency in the PDS is connected with the
keplerian frequency on the inner edge of the cold accretion disk. In this
case the decrease of the inner radius will result in the increase of the
characteristic frequency, the increase of the reflection amplitude and
(because of the feedback) the total softening of the primary spectrum (see Fig.5).
It is shown that the ``$\alpha - \Omega /2 \pi$'' dependence continues to
the soft state.

\bigskip

\bigskip

The full variant of the thesis can be found in \hfill

http://hea.iki.rssi.ru/$\sim$mikej/thesis/dis.html

\end{document}